IAPS | Institute for AI Policy and Strategy

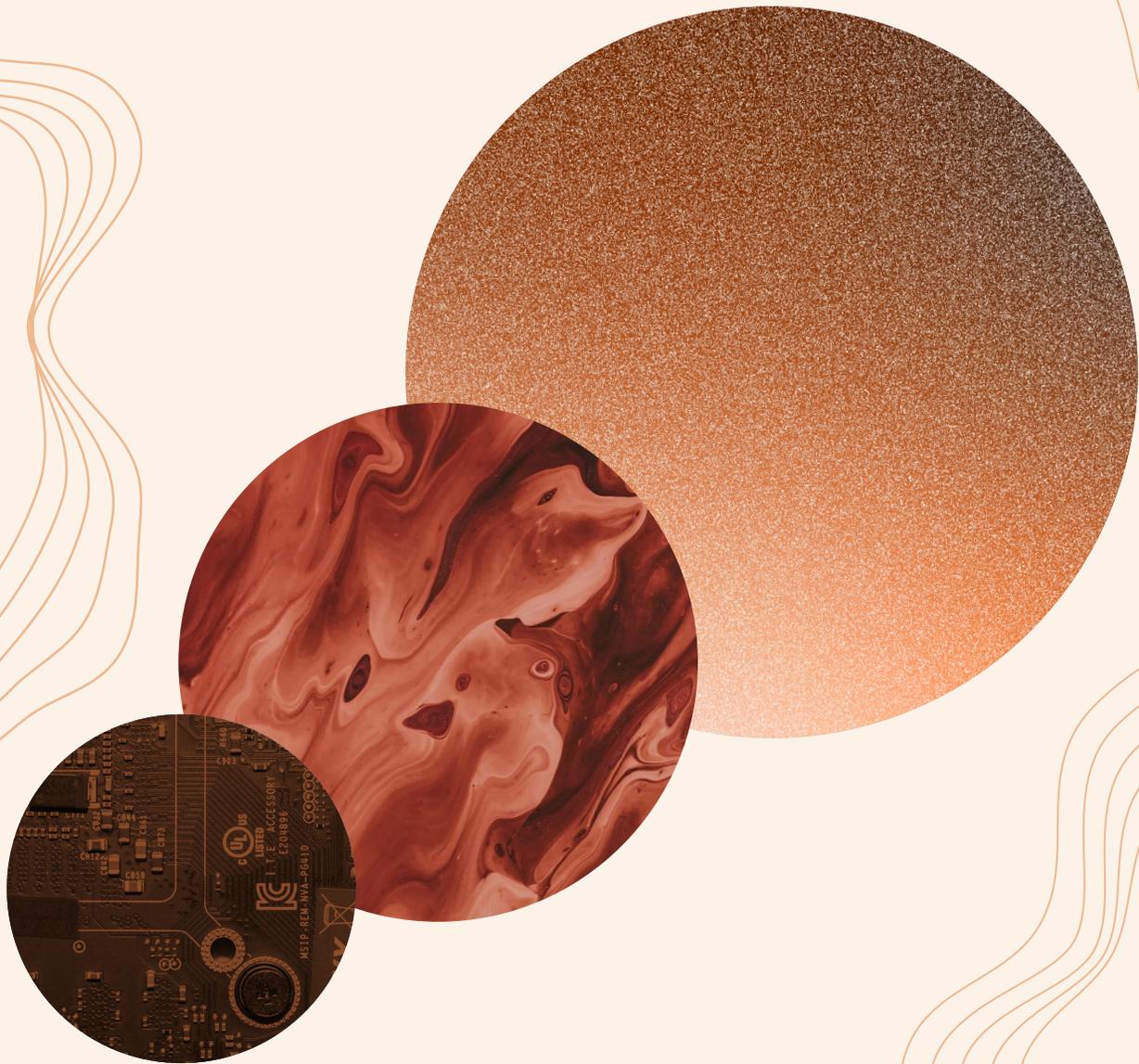

October 2024

# Understanding the First Wave of AI Safety Institutes

Characteristics, Functions, and Challenges

Renan Araujo, Kristina Fort, and Oliver Guest

# Executive Summary

In November 2023, the UK and US announced the creation of their AI Safety Institutes (AISIs). Five other jurisdictions have followed in establishing AISIs or similar institutions, with more likely to follow. While there is considerable variation between these institutions, there are also key similarities worth identifying.

**This report primarily describes one cluster of similar AISIs, the "first wave," consisting of the Japan, UK, and US AISIs**. Additionally, we compare the first wave to other AISI-like institutions in the EU, Canada, France, and Singapore.

First-wave AISIs have several fundamental characteristics in common:

- **Technical government institutions**.
- **Clear mandate related to the safety of advanced AI systems**. First-wave AISIs do not have "catch-all" responsibilities for AI within a jurisdiction.
- **No regulatory powers**.

**Safety evaluations are at the center of first-wave AISIs**. Safety evaluations are techniques that test AI systems across tasks to understand their behavior and capabilities on relevant risks, such as cyber, chemical, and biological misuse.

**First-wave AISIs have three core functions: research, standards, and cooperation**. These functions are critical to first-wave AISIs' work on safety evaluations, but also support other activities such as scientific consensus-building and foundational AI safety research.

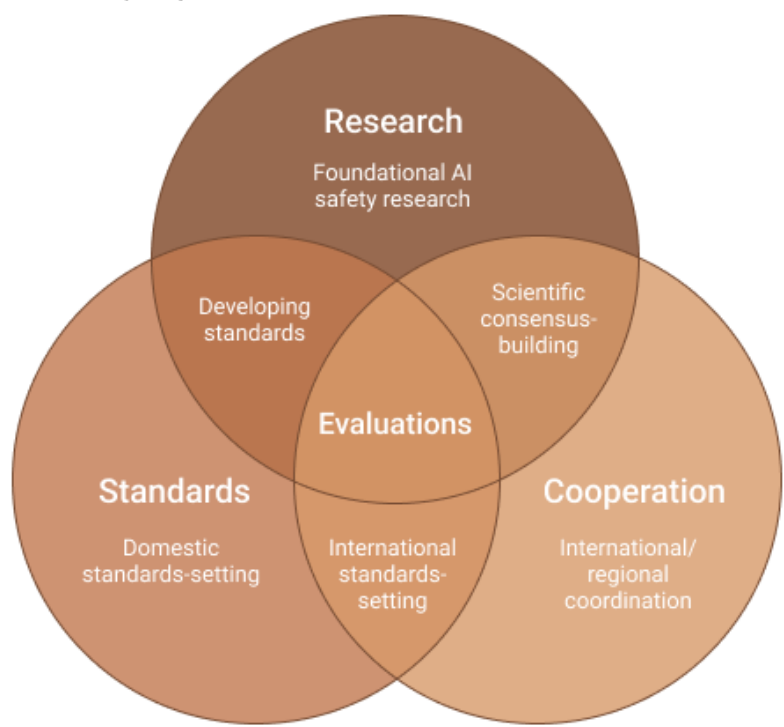

- **Research**: AISIs so far have emphasized research aimed at advancing the "science of AI safety" through technical work that is mainly empirical and problem-oriented, especially in the case of the UK and US AISIs. This has included research and testing around evaluations, foundational AI safety research, and efforts to build the field of AI safety as a whole.

- **Standards** are a more prescriptive function that AISIs play where they directly influence the practices of industry, government, and society through guidelines and protocols. AISIs' work on standards ranges from light-touch development of standardized procedures, such as in Japan and the US, to directly setting standards that feed into a wider regulatory process, such as in the EU.

- **Cooperation** is the role that AISIs play as a bridge between governments, industry, and society. AISIs also leverage their technical focus and state-backed legitimacy to boost international coordination on AI safety through initiatives like convenings and scientific consensus-building.

However, despite its growing popularity as an institutional reference, the AISI model is not free from **challenges and limitations**. Some analysts have criticized the first wave of AISIs for:

- **Specializing too much** on a sub-area and potentially neglecting concerns related to fields like national competitiveness and innovation, or fairness and bias.

- **Potential redundancies with existing institutions**, such as existing standards-developing bodies.

- **Their relationship with industry**, which has been productively close but might affect their impartiality.

Future developments may rapidly change this landscape, and particularities of individual AISIs may not be captured by our broad-strokes description. This policy brief aims to outline the core elements of first-wave AISIs as a way of encouraging and improving conversations on this novel institutional model, acknowledging this is just a simplified snapshot rather than a timeless prescription.

# Table of Contents

# 1 Introduction

In the last couple of years, rapid progress in AI systems has made AI policy an increasingly relevant priority for governments worldwide. To address the challenges associated with this novel policy field, a new institutional model has emerged: the AI Safety Institutes (AISIs).[1]

## 1.1 A brief overview of the current AISI landscape

**The emergence of AISIs is an interesting policy phenomenon.** It is one of the first—perhaps the first—institutional models of AI governance that different governments have adopted in a similarly shaped way. While explaining why AISIs have emerged is beyond the scope of this brief, we aim at outlining what is at the core of this new institutional model, what it means in practice, and what challenges lay ahead.

**There are currently seven jurisdictions with AISI-like institutions worldwide: the UK, the US, the EU, Japan, Singapore, France, and Canada**.[2] These are bodies that share some core characteristics and functions, even if they have different names or not entirely overlapping roles. In fact, they vary a lot among themselves, as show in Table 1 on the following page:

---

[1] Ziosi, Marta et al. "AISIs' Roles in Domestic and International Governance." Oxford Martin School, July 12, 2024. https://www.oxfordmartin.ox.ac.uk/publications/aisis-roles-in-domestic-and-international-governance.

[2] The Chinese "AI Safety Network" is sometimes also described as equivalent to an AISI. See, for example, Ziosi et al. 2024. However, the network appears to have relatively little government involvement. There is also little published information about the Network, making the Network difficult to characterize. It does not have a Chinese-language website. It is not clear that the projects described on the English-language website have input from the Network, rather than being work that participating institutions would have done anyway. "Chinese AI Safety Network." Accessed September 27, 2024. https://chinese-ai-safety.network/.

Table 1. Overview of AISIs[3]

| Jurisdiction | Activities | Funding | Institutional model |
|---|---|---|---|
| UK (AISI) | Evaluations, research, international coordination | £100 million until 2030 | First wave AISI |
| US (AISI) | Evaluations, developing and setting standards | $10 million for 2024/2025 | First wave AISI |
| Japan (AISI) | Evaluations, developing standards | Unclear | First wave AISI |
| EU (AI Office) | Codes of practice and regulation for AI, evaluations | €46.5 million, setup funding | Contemporaneous to the first wave, but distinct |
| Singapore (Digital Trust Center) | Research | $37 million, setup funding in 2022. Current funding unclear | Potentially similar to the first wave model |
| Canada (AISI) | Evaluations[4] | C$50 million[5] | Potentially similar to the first wave model |
| France (AI Evaluation Center) | Research and innovation, evaluations | Unclear | Potentially new trend |

Noticeably, **several institutions above are identified as 'AISIs', despite not being officially named as such**. While not uncommon for different governments to use distinct names and institutional frameworks to cover similar areas,[6] it is worth looking at the specific case of AISIs to justify these similarities. First, existing analyses often group these other AI-focused organizations together with organizations named 'AISI' because of their similar purposes and histories. Second, they are sometimes described by their own governments as AISIs.[7] Third, these institutions in practice play the role of AISIs on the international stage. For example, we expect Singapore's DTC and the EU AI Office to represent their jurisdictions during the inaugural meeting of the International Network of AI Safety Institutes in November.[8]

---

[3] Apart from Canada, information from Petropoulos, Alex. "The AI Safety Institute Network: Who, What and How?" ICFG (blog), September 10, 2024. https://icfg.eu/the-ai-safety-institute-network-who-what-and-how/.

[4] UK Government. "UK-Canada Science of AI Safety Partnership," May 20, 2024. https://www.gov.uk/government/publications/uk-canada-science-of-ai-safety-partnership/uk-canada-science-of-ai-safety-partnership.

[5] Canada, Department of Finance. "Remarks by the Deputy Prime Minister on Securing Canada's AI Advantage." Speeches, April 7, 2024. https://www.canada.ca/en/department-finance/news/2024/04/remarks-by-the-deputy-prime-minister-on-securing-canadas-ai-advantage.html.

[6] Using environmental matters as an example: Brazil has the Ministry of the Environment and Climate Change,, the US has the Environmental Protection Agency, and France has the Ministry for the Ecological Transition.

## 1.2 The first wave of AISIs

**The origins of AISIs can be traced back to November 2023, when the US and the UK announced the creation of their AISIs**.[9] The conception of the model is typically attributed to the UK, which announced a "Foundation Model Taskforce" in April 2023 to concentrate and catalyze government efforts around the most advanced models.[10] That taskforce was renamed the Frontier AI Taskforce in September 2023,[11] and later turned into the AI Safety Institute.[12] It has benefited from the growing staff and resourcing the UK government has started to pour into it: around £100 million and dozens of researchers, including some of the best AI safety experts in the world. Along with its AISI, the UK also launched a series of global summits on AI safety in November 2023. The two summits in this series so far have both been key events for the announcement of AISI-related news from various countries.[13]

---

[7] Infocomm Media Development Authority. "Digital Trust Centre Designated as Singapore's AI Safety Institute," May 22, 2024.
https://www.imda.gov.sg/resources/press-releases-factsheets-and-speeches/factsheets/2024/digital-trust-centre.

[8] Shepardson. "US to Convene Global AI Safety Summit in November." Reuters, September 18, 2024.
https://www.reuters.com/technology/artificial-intelligence/us-convene-global-ai-safety-summit-november-2024-09-18/.

[9] UK Government. "Introducing the AI Safety Institute," November 2023.
https://www.gov.uk/government/publications/ai-safety-institute-overview/introducing-the-ai-safety-institute.

[10] UK Government. "Initial £100 Million for Expert Taskforce to Help UK Build and Adopt next Generation of Safe AI," April 24, 2024.
https://www.gov.uk/government/news/initial-100-million-for-expert-taskforce-to-help-uk-build-and-adopt-next-generation-of-safe-ai.

[11] UK Government. "Frontier AI Taskforce: First Progress Report," September 7, 2023.
https://www.gov.uk/government/publications/frontier-ai-taskforce-first-progress-report/frontier-ai-taskforce-first-progress-report.

[12] UK Government. "AI Safety Institute: Third Progress Report," February 5, 2024.
https://www.gov.uk/government/publications/uk-ai-safety-institute-third-progress-report/ai-safety-institute-third-progress-report.

[13] UK Government. "AI Safety Summit," November 1, 2023. https://www.aisafetysummit.gov.uk/.

**The US AISI was also announced in November 2023, and has also been at the forefront of the AISI ecosystem**. It has $10 million in funding, around a dozen staff, and has gradually published several documents, such as its vision[14] and plan for global engagement on standard-setting.[15]

**Japan established an AISI in February 2024** under its Council for Science, Technology and Innovation, after having been a leader in the international governance of AI when it established the Hiroshima Process under its G7 presidency in 2023.[16]

> We identify the three AISIs set up by **the UK, the US, and Japan in 2023 and 2024 as the first wave of AISIs.**

These bodies were all heavily influenced by the process in the UK that culminated with the Bletchley summit, and share a set of common characteristics and functions that revolve around safety evaluations of advanced AI systems—which we explore in more detail in the following sections.

## 1.3 Other relevant institutions and possible waves

**Alongside the first wave of AISIs, the EU AI Office is a relevant contemporaneous institutional development** we will explore. The EU was an early starter in designing an institution for governing AI in 2021,[17] and launched its EU AI Office in early 2024 after the enactment of the EU AI Act. The EU AI Office has a much broader and stronger mandate than first-wave AISIs: it regulates the whole AI market in the European Union. At the same time, it plays roles equivalent to that of an AISI in many ways, such as having a unit dedicated to AI Safety (Unit 3),[18] emphasizing evaluations of general-purpose AI models as a core activity,[19] establishing technical direct partnerships with AISIs,[20] and representing the EU in AISI-related events, including the international network of AISIs.[21]

---

[14] US AI Safety Institute. "Strategic Vision." NIST, May 20, 2024. https://www.nist.gov/aisi/strategic-vision.

[15] Dunietz, Jesse, Elham Tabassi, Mark Latonero, and Kamie Roberts. "A Plan for Global Engagement on AI Standards." NIST, July 26, 2024. https://www.nist.gov/publications/plan-global-engagement-ai-standards.

[16] Habuka, Hiroki, and David U. Socol de la Osa. "Shaping Global AI Governance: Enhancements and Next Steps for the G7 Hiroshima AI Process." Center for Strategic and International Studies, May 24, 2024. https://www.csis.org/analysis/shaping-global-ai-governance-enhancements-and-next-steps-g7-hiroshima-ai-process.

[17] European Parliament. "EU AI Act: First Regulation on Artificial Intelligence," August 6, 2023. https://www.europarl.europa.eu/topics/en/article/20230601STO93804/eu-ai-act-first-regulation-on-artificial-intelligence.

**The other jurisdictions that have established AISIs are Singapore, France, and Canada.** However, these organizations are either distinct from the first wave model or there is not enough information about them yet to analyze their core functions and structure.

**Singapore repurposed an existing government body to focus more on safety and evaluations in the model of first-wave AISIs**,[22] but there is not enough information to determine whether it has effectively become a similar body in terms of characteristics and functions.

**Canada has started what seems similar to first-wave-AISIs**,[23] but there have not been public updates about it yet.

**The case of France is distinctive because it has prioritized elements other than safety in its AI governance approach, particularly innovation**. These additional priorities have been

---

[18] European Commission. "European AI Office | Shaping Europe's Digital Future," 2024. https://digital-strategy.ec.europa.eu/en/policies/ai-office.

[19] Idem.

[20] European Commission. "U.S. AI Safety Institute and European AI Office Technical Dialogue | Shaping Europe's Digital Future," July 12, 2024. https://digital-strategy.ec.europa.eu/en/news/us-ai-safety-institute-and-european-ai-office-technical-dialogue, Mokander, Jakob, Camilla Bougrine, Kevin Zandermann, Basma Abassi, and Zach Meyers. "Exploring EU-UK Collaboration on AI: A Strategic Agenda." Tony Blair Institute for Global Change, July 31, 2024. https://institute.global/insights/tech-and-digitalisation/exploring-eu-uk-collaboration-on-ai-a-strategic-agenda.

[21] US Department of Commerce. "U.S. Secretary of Commerce Raimondo and U.S. Secretary of State Blinken Announce Inaugural Convening of International Network of AI Safety Institutes in San Francisco." U.S. Department of Commerce, September 18, 2024. https://www.commerce.gov/news/press-releases/2024/09/us-secretary-commerce-raimondo-and-us-secretary-state-blinken-announce.

[22] Infocomm Media Development Authority. "Digital Trust Centre Designated as Singapore's AI Safety Institute," May 22, 2024. https://www.imda.gov.sg/resources/press-releases-factsheets-and-speeches/factsheets/2024/digital-trust-centre.

[23] UK Government. "UK-Canada Science of AI Safety Partnership," May 20, 2024. https://www.gov.uk/government/publications/uk-canada-science-of-ai-safety-partnership/uk-canada-science-of-ai-safety-partnership.

embodied by its remodeling of the AI Safety Summit to become their AI Action Summit.[24] Even though the French agency seems centered around evaluations similar to first-wave AISIs (and is labeled the AI Evaluation Center), due to this increased focus on innovation and action it might be the case that France ushers in a second wave of AISIs in its 2025 summit.

It is also worth noting that **a Chinese AI Safety Network has been launched**, bringing together several AI safety-relevant Chinese organizations.[25] However, **it lacks strong government links** and there has not been public news about its actions so far.

**A potential second wave of AISIs with more focus on innovation or AI benefits alongside safety could potentially be appealing to a wider set of nations, particularly developing countries**. This could reflect the state of discussions in India, for example, where the implications of AI on fields such as health, education, and agriculture is top of mind for local stakeholders that simultaneously acknowledge the relevance of safety.[26]

However, as of August 2024, there is not enough information for us to productively analyze a potential second wave of AISIs—so we will turn to the first wave for the moment.

---

[24] Élysée Palace. "AI Action Summit." elysee.fr, July 10, 2024. https://www.elysee.fr/en/ai-action-summit.

[25] Chinese AI Safety Network. "Chinese AI Safety Network." Accessed September 27, 2024. https://chinese-ai-safety.network/.

[26] Chaudhuri, Rudra. "Disrupting AI Safety Institutes: The India Way." Carnegie Endowment for International Peace, September 17, 2024. https://carnegieendowment.org/posts/2024/09/disrupting-ai-safety-institutes-the-india-way?lang=en.

# 2 Fundamental characteristics

**First-wave AI Safety Institutes have been safety-focused, governmental, technical institutions** intended to evaluate and contribute to the safety of advanced AI systems.

## 2.1 Safety-focused

Considering AISIs originated in the context of the Bletchley AI Safety Summit organized by the UK government, the focus on AI safety is naturally a key defining characteristic. The Bletchley Declaration, signed by all jurisdictions that participated in the Summit, states that "**AI should be designed, developed, deployed, and used, in a manner that is safe, in such a way as to be human-centric, trustworthy and responsible**."[27] In the context of the Bletchley declaration, safety is related to the significant risks posed by the "frontier" of AI:[28]

> "Particular safety risks arise at the 'frontier' of AI, understood as being those highly capable general-purpose AI models, including foundation models, that could perform a wide variety of tasks - as well as relevant specific narrow AI that could exhibit capabilities that cause harm - which match or exceed the capabilities present in today's most advanced models. Substantial risks may arise from potential intentional misuse or unintended issues of control relating to alignment with human intent. These issues are in part because those capabilities are not fully understood and are therefore hard to predict. We are especially concerned by such risks in domains such as cybersecurity and biotechnology, as well as where frontier AI systems may amplify risks such as disinformation."

One reason why focusing on this definition of safety is relevant is because AI development will have cross-sectoral implications that will be better tackled by sectoral government agencies. Risk management and mitigation involving advanced models, on the other hand, still benefits from a dedicated body considering how quickly the most advanced models evolve, and how much work governments need to do to catch up with the frontier and keep potentially harmful outcomes at bay.

---

[27] UK Government. "The Bletchley Declaration by Countries Attending the AI Safety Summit, 1-2 November 2023," November 1, 2023.
https://www.gov.uk/government/publications/ai-safety-summit-2023-the-bletchley-declaration/the-bletchley-declaration-by-countries-attending-the-ai-safety-summit-1-2-november-2023.

[28] Ibid.

## 2.2 Governmental

**AISIs in the first wave have been governmental institutions**, which has provided them with the **authority, legitimacy, and resources (funding and talent)** necessary to address AI safety on a national and global scale.

Table 2. Placement of first-wave AISIs within each government

| Jurisdiction | Agency | Department/Ministry |
|---|---|---|
| UK[29] | n/a | Department for Science, Innovation and Technology (DSIT) |
| US[30] | National Institute of Standards and Technology (NIST) | Department of Commerce |
| Japan[31] | Information-technology Promotion Agency (IPA) | Ministry of Economy, Trade and Industry |

For example, AISIs' governmental status has likely increased their ability to negotiate access to leading models in order to run evaluations on them; companies face a race of stronger enforcement mechanisms from the government if they do not voluntarily comply. AISIs may also have benefitted from having access to sensitive information from other government agencies that would not be otherwise available to non-governmental organizations.[32]

Being part of the government has also contributed to their agendas being more strongly guided by public interest. This may have contributed to minimizing the risk of capture by particular interest groups compared to, for example, industry-led setups—even though the relationship between AISIs and the private sector remains a challenge, as we explore later in this brief. Still, being part of government has helped AISIs to work inclusively with various stakeholders across civil society, academia, industry, other agencies, and foreign governments.

---

[29] UK Government. "About the AI Safety Institute | AISI." Accessed September 27, 2024. https://www.aisi.gov.uk/about.

[30] US AI Safety Institute. "U.S. Artificial Intelligence Safety Institute." NIST, October 26, 2023. https://www.nist.gov/aisi.

[31] Japan AI Safety Institute. "AI Safety Institute," 2024. https://aisi.go.jp/wp-content/uploads/2024/07/20240627_AboutAISI_en.pdf.

[32] This may have been particularly relevant for AISIs in the US and the UK, which more clearly have lacked regulatory powers, though might have been less essential for the EU AI Office, which already has enforcement mechanisms.

## 2.3 Technical

**The third fundamental characteristic of AISIs is that they have a focus on technical expertise**. First-wave AISIs have been centered around technical professionals and have had at their core a focus on the substance of AI safety. Having such expertise has been connected to the pursuit of first-wave AISIs of an evidence-based approach to AI safety. This focus is connected with their goals to attract top talent, move quickly with minimal political friction, and collaborate more widely with other centers of excellence domestically and internationally. Some AISIs have articulated this as an emphasis on "pushing the frontier of AI safety."[33]

This focus on technical frontier knowledge may have contributed to the ability of some AISIs to **attract top talent** from industry, academia, and civil society, besides government officials. In particular, AISIs in the US and the UK have recruited some of the most influential AI researchers, such as Paul Christiano,[34] who pioneered reinforcement learning with human feedback, and Geoffrey Irving,[35] a key contributor to safety areas such as 'AI safety via debate' and interpretability.

With this combination of technical-oriented features, **AISIs may gradually become government powerhouses that help governments to stay on top of fast-paced AI developments**. That said, governments with AISIs often have other units with advanced technical expertise relating to AI and AI safety. Perhaps the clearest example of this is the "Safeguarded AI" program within the UK's Advanced Research + Innovation Agency.[36]

## 2.4 What first-wave AISIs are not

**Defining first-wave AISIs also comes with the challenge of outlining what institutions are not in this category**. Achieving such a definition is difficult because AISIs are a very new organizational model. However, it is useful to concretely define AISIs, or at least specific kinds of

---

[33] US AI Safety Institute. "Strategic Vision." NIST, May 20, 2024. https://www.nist.gov/aisi/strategic-vision.

[34] NIST. "U.S. Commerce Secretary Gina Raimondo Announces Expansion of U.S. AI Safety Institute Leadership Team," April 16, 2024.
https://www.nist.gov/news-events/news/2024/04/us-commerce-secretary-gina-raimondo-announces-expansion-us-ai-safety.

[35] UK Government. "AI Safety Institute: Third Progress Report," February 5, 2024.
https://www.gov.uk/government/publications/uk-ai-safety-institute-third-progress-report/ai-safety-institute-third-progress-report.

[36] Aria. "Safeguarded AI." Accessed July 9, 2024. https://www.aria.org.uk/programme-safeguarded-ai/.

AISIs, for several reasons: governments may want to set up AISIs and may be confused by the different models, the emerging network of AISIs may have a hard time determining membership criteria (and non-members may feel unfairly excluded), and it is generally useful to pinpoint what is distinctive about (first-wave) AISIs in order to more precisely discuss their strengths and weaknesses as an institutional model.

We describe here **three broad characteristics we see as relatively incompatible with the first wave of AISIs** and our opinion about the implications of steering clear of these traits:

- Catch-all-AI responsibilities
- Complex bureaucracies
- Regulatory powers

First, **AISIs so far have not been a catch-all organization for everything AI-related the government aims to do**. The focus on safety has been crucial because AI as a policy field is already large in scope and will only continue to grow as the technology becomes more transformative across sectors and is more widely adopted across more industries. Even leaving aside sector-specific governance for each sectoral government agency, the broad responsibility of setting AI policy is likely better suited to a constellation of government agencies or a high-level ministry or department rather than an AISI. The edge AISIs have is the focus on a time-sensitive problem that will have downstream effects on all other AI-related issues: ensuring society has the knowledge and tools to make advanced models safe.

Secondly, **AISIs have benefitted from being nimble organizations instead of complex bureaucracies**. Being highly-specialized has been necessary for AISIs to help the government stay up to speed with advanced AI, a field that changes much more quickly than the typical policy area. AISIs in the first wave have benefitted from being relatively small, filled with experts, mission-driven, and free of complicated bureaucratic procedures. This is also why isolating AISIs from traditional government responsibilities has been important: bureaucracy becomes more relevant the closer an agency is to activities such as deciding budget allocations, welfare distribution, and political appointments because ensuring transparency and fairness are respected is paramount. By focusing on a few technical functions, AISIs have been protected from the burdensome responsibilities that come with such powers.

Third, **first-wave AISIs have typically not been granted hard regulatory powers so far**. Besides avoiding the bureaucratic burdens that come with stronger responsibilities, avoiding regulatory powers has also minimized political complexities that come with such an economically and socially relevant field as AI.

# 3 Core functions

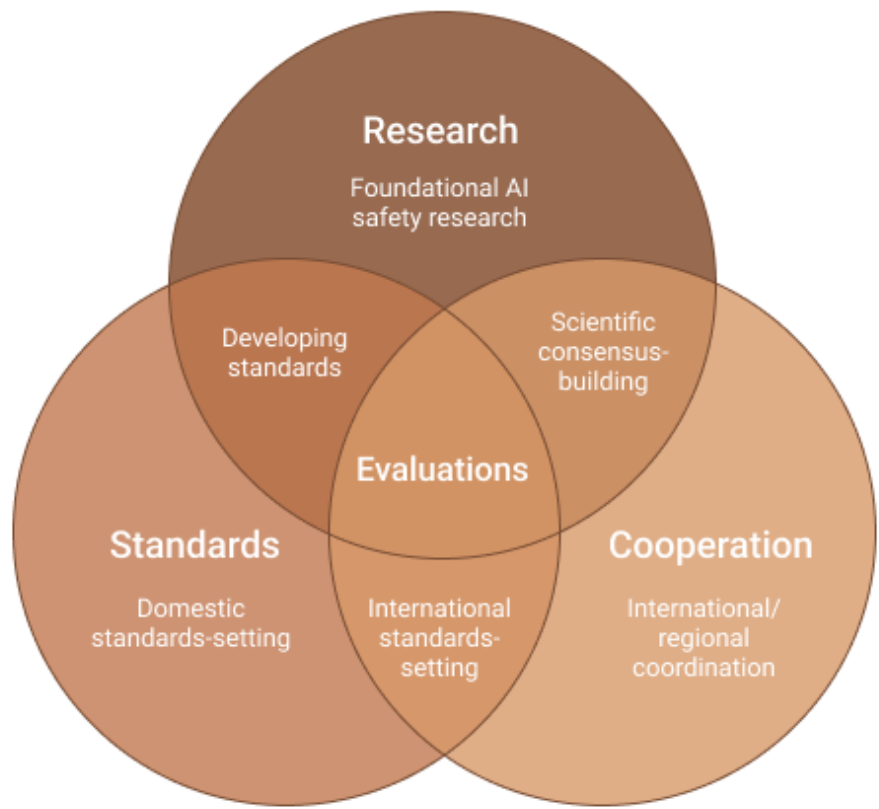

Figure 1. Diagram of core functions of the first wave of AISIs

There are many granular activities that AISIs conduct or aspire to perform, with some relevant variation across jurisdictions.[37] In this brief, we identify three functions that the first wave of AISIs have noted as top priorities and that contribute to their institutional identity: research, standards, and cooperation. In each of them, however, we notice the pattern of a core cross-functional activity in evaluations.

---

[37] Ziosi, Marta, Claire Dennis, Robert Trager, Simeon Campos, Ben Bucknall, Charles Martinet, Adam Smith, and Merlin Stein. "AISIs' Roles in Domestic and International Governance." Oxford Martin School, July 12, 2024. https://www.oxfordmartin.ox.ac.uk/publications/aisis-roles-in-domestic-and-international-governance and Variengien, Alexandre, and Charles Martinet. "AI Safety Institutes: Can Countries Meet the Challenge?" OECD.AI Policy Observatory, July 29, 2024. https://oecd.ai/en/wonk/ai-safety-institutes-challenge.

## 3.1 Evaluations: the core cross-functional activity

AI evaluations are techniques and procedures that test AI systems across tasks to understand their behavior and capabilities.[38] **Safety evaluations test particular AI capabilities that present relevant risks**, including cyber, chemical, biological misuse, autonomous capabilities, and the effectiveness of safeguards.[39]

**At their core, AISIs in the first wave have been about evaluations**. A considerable portion of AISIs' research so far has been dedicated to evaluations, which can also be used as benchmarks for the development of standards.[40] Since the research done by AISIs has been empirical and decision-oriented, evaluations are a particularly promising avenue as they provide tests that, when passed or not, should lead to predefined actions. This practical usefulness has also made evaluations useful for establishing cooperation agreements, as it creates a shared language for decision-makers on AI safety.

In that context, **AISIs have played a pivotal role in growing and strengthening the field of safety evaluations**. The field started to take off within industry and a few nonprofit research organizations in late 2022 and early 2023,[41] but being restricted to independent evaluators posed a limit on the field's ability to keep up with industry's pace. With AISIs, by leveraging the expertise of these researchers through partnerships and adding teeth to the requirements for companies to provide access to their models,[42] the field of AI safety has gradually grown and become more established.

Here is some of the work various AISIs have done on developing evaluations:

---

[38] NIST. "AI Test, Evaluation, Validation and Verification (TEVV)." 2024. https://www.nist.gov/ai-test-evaluation-validation-and-verification-tevv.

[39] UK AI Safety Institute. "AISI Research | Developing and conducting model evaluations. Advancing foundational safety and societal resilience research." Accessed 30 September 2024. https://www.aisi.gov.uk/category/research.

[40] Ibid.

[41] METR. "2023 Year In Review," February 7, 2024. https://metr.org/blog/2024-02-07-2023-year-in-review/ and Apollo Research. "The first year of Apollo Research," May 29, 2024. https://www.apolloresearch.ai/blog/the-first-year-of-apollo-research.

[42] UK Government. "Leading frontier AI companies publish safety policies," October 27, 2023. https://www.gov.uk/government/news/leading-frontier-ai-companies-publish-safety-policies.

Table 3. Work or plans by AISIs on evaluations

| Jurisdiction | Work or plans on evaluations |
|---|---|
| UK | Evaluations have covered:[43]<ul><li>"Whether the models could potentially be used to facilitate cyber-attacks;</li><li>"Whether they could provide expert-level knowledge in chemistry and biology that could be used for positive but also harmful purposes;</li><li>"Whether they were capable of autonomously taking sequences of actions (operating as 'agents') in ways that might be difficult for humans to control and</li><li>"Whether they were vulnerable to 'jailbreaks' or users attempting to bypass safeguards to elicit potentially harmful outputs (e.g. illegal or toxic content)."</li></ul>Additionally, UK AISI open-sourced their testing framework, Inspect.[44] |
| US | Plans include:[45]e<ul><li>"AISI will champion the development of empirically grounded tests, benchmarks, and evaluations of AI models, systems, and agents to find practical solutions for both near and long-term AI safety challenges. Accordingly, we aim to launch a range of projects under this goal, some of which may seek to:<ul><li>"Perform and coordinate technical research to improve or create needed safety guidelines and technical safety tools and techniques (...)</li><li>"Conduct pre-deployment TEVV of advanced models, systems, and agents to assess potential and emerging risks (...)</li><li>"Conduct TEVV of advanced AI models, systems, and agents to develop scientific understanding and documentation of the range of existing risks."</li></ul></li></ul>(Note the US also has the Assessing Risks and Impacts of AI program under NIST, which also plans to do model evaluations but is a separate entity from AISI.)[46] |
| Japan | Its role description includes:[47]<ul><li>"AISI supports the government by conducting surveys on AI safety, examining evaluation methods, and creating standards."</li></ul> |

As shown by the overview above, **all first-wave AISIs have announced plans to do evaluations-related work**. However, only the UK has published significant work on concrete

---

evaluations.[48] The US has the most detailed set of plans, and there is some evidence it has started work related to evaluations,[49] but no publication has provided a rundown of their work on evaluations so far. **The UK and the US are likely to be upstream in the evaluations pipeline, contributing to the actual test and benchmark development process**.

On the other hand, **Japan seems more likely to be downstream in the evaluations pipeline**. The Japan AISI is planning to incorporate research and evaluations done by other AISIs through international partnerships, and has explicitly announced its AISI will not conduct R&D. This might also be the case for other jurisdictions, like the EU and Canada. Their bodies seem unlikely to have plans to be at the frontier of evaluations work apart from partnering with the UK and the US, considering their level of resourcing, talent, and access to top AI companies.

As we will explore in more detail in the following sections, each of the other core functions manifests itself in ways that are deeply connected to evaluations:

- **Research**: to advance the science of AI safety, AISIs investigate model capabilities that present threats in a replicable way, building the foundation for how to manage them
- **Standards**: evaluations are the stepping stone to standardized protocols and benchmarks that contribute to standards-developing processes
- **Cooperation**: building a coordinated body of knowledge for information-sharing on models and safety measures and techniques

Each of these functions go beyond evaluations as well, as shown by the diagram above. However, evaluations are always either at their core or a relevant activity.

## 3.2 Research

Another core function of first-wave AISIs is research for the advancement of the "science of AI safety"[50] by investing in new technical knowledge and tools to improve our understanding of how to make advanced AI systems safer. **The research that first-wave AISIs have conducted has**

---

[48] UK AI Safety Institute. "Advanced AI Evaluations at AISI: May Update | AISI," May 20, 2024. https://www.aisi.gov.uk/work/advanced-ai-evaluations-may-update.

[49] Anthropic. "Introducing Claude 3.5 Sonnet," June 21, 2024. https://www.anthropic.com/news/claude-3-5-sonnet.

[50] US AI Safety Institute. "Strategic Vision." NIST, May 20, 2024. https://www.nist.gov/aisi/strategic-vision.

**been mainly empirical and action-relevant**, aiming to be actionable by the government, companies, and other stakeholders working on the safety of advanced AI.

### a. Developing the science of AI safety

Most AISIs, particularly in the US and UK, have set themselves **the goal of contributing to AI safety from a technical and empirical perspective**.[51] This primarily means having in-house expertise to perform research themselves, but also includes funding programs and partnering with industry, academia, and civil society to advance AI safety research.

**Research has been a core function for first-wave AISIs because AI safety is a new but growing field.** Little is known right now about what risks AI systems pose, and governments have a hard time ensuring the public interest stays up-to-speed with this quickly-evolving technology. While developments in the capabilities of large language models have grown quickly, the behavior of advanced AI systems remains challenging to explain and steer.[52] While industry, academia, and civil society have contributed to advancing the field, AISIs aim at strengthening these efforts through the uniquely public interest focus that governments can provide. Additionally, by becoming government powerhouses for AI expertise, AISIs may also help governments to better understand the frontier of AI development, positioning themselves to manage risks and opportunities.

**In the two AISIs at the forefront of research work, the US and the UK, research agendas are primarily empirical and problem-oriented rather than theoretical or abstract.** The US AISI describes its research as focusing on:

> "Perform and coordinate technical research to improve or create needed safety guidelines and technical safety tools and techniques, such as techniques for detection of synthetic content, best practices for model security, and technical safeguards and mitigations at the level of models, systems, and agents. These projects may involve both foundational and applied research. For its applied research projects, AISI intends to leverage in-house and external foundational research, as well as existing guidelines, methods, and standards."[53]

---

[51] See NIST. "Strategic Vision." *NIST*, May 20, 2024. https://www.nist.gov/aisi/strategic-vision and UK AI Safety Institute. "About the AI Safety Institute | AISI," 2024. https://www.aisi.gov.uk/about.

[52] Owen, David. "How Predictable Is Language Model Benchmark Performance?" Epoch AI, June 9, 2023. https://epochai.org/blog/how-predictable-is-language-model-benchmark-performance.

[53] US AI Safety Institute. "Strategic Vision." NIST, May 20, 2024. https://www.nist.gov/aisi/strategic-vision, p. 4.

One example of output from the US has been their initial draft on managing misuse risk for dual-use foundation models.[54] This publication is much closer to the applied end of research, bordering policy recommendations in the style of standards-developing work, true to the actionable spirit the US AISI ascribes itself in its vision document. The piece provides guidelines for improving the safety, security, and trustworthiness of AI systems, such as listing key threat profiles for misuse risk, how to define risk thresholds, and measures against model theft.

The UK AISI's research page describes its mission similarly:

> "AISI develops and conducts model evaluations to assess risks from cyber, chemical, biological misuse; autonomous capabilities and the effectiveness of safeguards. We also are also working to advance foundational safety and societal resilience research."[55]

An example of research by UK AISI is their exploration of safety cases.[56] A safety case is a "structured argument that an AI system is safe within a particular training or deployment context." Safety cases work builds on agendas such as interpretability-based evaluations, AI control[57] and associated threat models, and automated AI safety, including through partnerships with organizations specialized in fields like these, to lay the groundwork for effectively mapping the problem space around evaluating model safety.

## b. Building the field of AI safety

**AISIs are approaching research in a fairly collaborative way, which contributes to building the research field of AI safety**. Besides performing research themselves, as outlined above, AISIs are partnering with industry, academia, civil society, and other governments. Such partnerships have taken different forms, ranging from formal memoranda of understanding to open calls for proposals and grant programs.

---

[54] US AI Safety Institute. "Managing Misuse Risk for Dual-Use Foundation Models." Initial Public Draft, July 2024. https://doi.org/10.6028/NIST.AI.800-1.ipd.

[55] UK AI Safety Institute. "Research," 2024. https://www.aisi.gov.uk/category/research.

[56] Irving, Geoffrey. "Safety Cases at AISI," August 2, 2024. https://www.aisi.gov.uk/work/safety-cases-at-aisi.

[57] Greenblatt, Ryan, Buck Shlegeris, Kshitij Sachan, and Fabien Roger. "AI Control: Improving Safety Despite Intentional Subversion." arXiv, July 23, 2024. https://doi.org/10.48550/arXiv.2312.06942.

**Among AISIs, there have been several cooperation agreements**. The establishment of an international network of AISIs has been a notable development towards cooperation.[58] Bilateral agreements have also been numerous, mainly among subsets of the EU, UK, US, Canada, and Japan.[59] These partnerships among AISIs are aimed at facilitating the exchange of information ranging from new technical approaches to testing, evaluations, and verification strategies to sharing information collected from AI companies, including model access.[60]

Along the lines of facilitating the exchange of information from companies, **there have also been partnerships between AISIs and industry**. The US AISI has signed agreements with Anthropic and OpenAI on AI safety research, testing, and evaluation.[61] Anthropic also engaged with the UK AISI before deploying Claude 3.5 Sonnet for a safety evaluation that was also shared with the US AISI.[62] Other types of collaborative engagement have included calls for comments and feedback[63] and grant programs.[64]

**However, there are also challenges to these collaborations**.[65] It is in the interests of AISIs to have access to industry information about the latest models in a timely manner, but companies may find this additional scrutiny burdensome and fail to collaborate—which has already happened with UK AISI attempts, according to some journalistic pieces denied by UK AISI itself.[66] This issue becomes more complex with the inter-AISI collaboration agreements—the network of AISIs hopes to facilitate information sharing but **companies might not want to have sensitive model data shared freely across multiple jurisdictions**.[67] While in some ways an AISI network contributes to lowering the compliance burden on companies by centralizing information exchange, it also increases the vulnerability surface of potentially dual-use, proprietary information.

---

[58] Shepardson. "US to Convene Global AI Safety Summit in November." *Reuters*, September 18, 2024. https://www.reuters.com/technology/artificial-intelligence/us-convene-global-ai-safety-summit-november-2024-09-18/.

[59] UK Government. "UK-Canada Science of AI Safety Partnership," May 20, 2024. https://www.gov.uk/government/publications/uk-canada-science-of-ai-safety-partnership/uk-canada-science-of-ai-safety-partnership.

[60] UK AI Safety Institute. "Governance | AISI," 2024. https://www.aisi.gov.uk/category/governance.

[61] NIST. "U.S. AI Safety Institute Signs Agreements Regarding AI Safety Research, Testing and Evaluation With Anthropic and OpenAI," August 29, 2024. https://www.nist.gov/news-events/news/2024/08/us-ai-safety-institute-signs-agreements-regarding-ai-safety-research.

[62] Anthropic. "Expanding access to Claude for government," June 26, 2024. https://www.anthropic.com/news/expanding-access-to-claude-for-government.

[63] NIST. "NIST Calls for Information to Support Safe, Secure and Trustworthy Development and Use of Artificial Intelligence," December 19, 2023. https://www.nist.gov/news-events/news/2023/12/nist-calls-information-support-safe-secure-and-trustworthy-development-and.

[64] UK AI Safety Institute. "Systemic AI Safety Fast Grants," 2024. https://www.aisi.gov.uk/grants.

[65] As we explore in more detail in Challenges.

In short, **research is a key function of AISIs as they connect the frontier of AI safety to decision-relevant recommendations across government, industry, and civil society**. In the following section, we will analyze evaluations and standards—a different function that has some overlap with research, but points to a distinct, more prescriptive, direction.

## 3.3 Standards

**In this function, AISIs' work turns towards a more prescriptive role** than research: it aims at influencing how various stakeholders, particularly industry and other governments, will approach AI safety. Standards work done by AISIs varies in strength depending on the jurisdiction, ranging from mere examination of evaluations and light-touch standards-developing practices in Japan to setting and enforcing standards as part of a wider regulatory process in the EU.

**Typically, standards are developed, set, and maintained by Standards Developing Organizations (SDOs)**. These organizations might have a general standard-related scope, like the International Organization for Standardization (ISO), or a sector-specific role, like the International Electrotechnical Commission (IEC), the Institute of Electrical and Electronics Engineers (IEEE), and the International Telecommunication Union (ITU). Others have a regional scope, such as the European Committee for Standardization (CEN) and the European Electrotechnical Committee for Standardization (CENELEC) for the EU or other national standards bodies.[68] AISIs would likely feed into all these existing pipelines.

---

[66] Manancourt, Vincent. "Rishi Sunak Promised to Make AI Safe. Big Tech's Not Playing Ball." Politico, April 26, 2024. https://www.politico.eu/article/rishi-sunak-ai-testing-tech-ai-safety-institute/.

[67] US Department of Commerce. "U.S. Secretary of Commerce Raimondo and U.S. Secretary of State Blinken Announce Inaugural Convening of International Network of AI Safety Institutes in San Francisco." U.S. Department of Commerce, September 18, 2024.
https://www.commerce.gov/news/press-releases/2024/09/us-secretary-commerce-raimondo-and-us-secretary-state-blinken-announce.

[68] Christopher Thomas and Florian Ostmann. "Enabling AI Governance and Innovation through Standards." UNESCO, March 15, 2024.
https://www.unesco.org/en/articles/enabling-ai-governance-and-innovation-through-standards.

**The main contribution from AISIs would arguably be speed and resources to ensure a high-quality standards-developing and setting process**. Standards organizations are often slow as they need to ensure that the standardization process is inclusive, transparent, and effectively leads to standards that contribute to the specific field rather than hasty results that might hinder innovation or interoperability. By bringing in AI safety-specific top expertise and funding, AISIs help ensure the standard-setting process is higher quality and able to move more quickly than it would by default in other bodies. On the other hand, challenges are still to be expected on the quality and fidelity of these processes as standards are highly consequential and AISIs are very new institutions.[69]

In this brief, for a more convenient analysis, we look at the standardization process in two stages: the development and the setting stages. Development is the process through which organizations identify the best practices and formulate standardized guidelines. In the standards-setting stage, organizations help turn guidelines into actionable, often enforceable rules.

## a. Developing standards

Considering how novel the field of AI safety is and how empirical and decision-relevant AISIs aspire to be, **there is often a strong overlap between the work developing standards and research, especially on evaluations**.

**Standards "provide common and repeatable rules, guidelines and characteristics for activities or their results,"[70]** which is similar to the standardized nature that safety evaluations gradually developed. After all, as previously mentioned, evaluations provide tests whose results lead to predefined actions depending on whether they are passed or not.

For example, the NIST Plan for Global Engagement on AI Standards, when describing "measurement methods and metrics" under the topics that are urgently needed and ready for standardization (emphasis added):[71]

---

[69] Fort, Kristina. "The Role of AI Safety Institutes in Contributing to International Standards for Frontier AI Safety." arXiv.org, September 17, 2024. https://arxiv.org/abs/2409.11314v1.

[70] Christopher Thomas and Florian Ostmann. "Enabling AI Governance and Innovation through Standards." UNESCO, March 15, 2024. https://www.unesco.org/en/articles/enabling-ai-governance-and-innovation-through-standards.

[71] NIST. "A Plan for Global Engagement on AI Standards." 2024. https://doi.org/10.6028/NIST.AI.100-5, pp. 10-11.

"Measurement methods and metrics. Shared testing, evaluation, verification, and validation (TEVV) practices for AI models and systems would open the way for more rigorous discussions about capabilities, limitations, risks, benefits, appropriate or inappropriate use, AI assurance, and more. In particular, **TEVV standards would define the performance metrics for performance-based standards, which in turn allow defining what constitutes an effective risk mitigation**. (...) Further work may be needed before it is possible to standardize some testing and evaluation protocols (e.g., AI red-teaming and evaluating impacts on people in realistic settings), some types of measurements (e.g., measuring effectiveness and robustness of interventions to mitigate risks), and what TEVV approaches are most effective for some types of risks (e.g., risks to safety or information integrity), all of which would merit accelerated study."

**For examples of research work that has the potential to feed the development of standards, we can turn to the UK AISI**. In the description of their approach to evaluations, UK AISI acknowledges the lack of established standards for best practices and sets itself the goal to "build an evaluations process for assessing the capabilities of the next generation of advanced AI systems."[72] Since then, they have been using standardized benchmarks for testing model behavior[73] and have open sourced their own evaluation framework,[74] both of which contribute to improving the quality of standards in the field.

The work of other institutions that are not conducting research or directly setting standards may also fall under this category. **The Japanese AISI plans to conduct surveys on AI safety, examine evaluation methods, and create standards**. Goals of that work include strengthening the local business environment and being part of the international network contributing to AI safety.[75]

---

[72] UK Government. "AI Safety Institute Approach to Evaluations," February 9, 2024. https://www.gov.uk/government/publications/ai-safety-institute-approach-to-evaluations/ai-safety-institute-approach-to-evaluations.

[73] Mazeika, Mantas, Long Phan, Xuwang Yin, Andy Zou, Zifan Wang, Norman Mu, Elham Sakhaee, et al. "HarmBench: A Standardized Evaluation Framework for Automated Red Teaming and Robust Refusal." arXiv.org, February 6, 2024. https://arxiv.org/abs/2402.04249v2 quoted in UK Government. "Advanced AI Evaluations at AISI: May Update | AISI," May 20, 2024. https://www.aisi.gov.uk/work/advanced-ai-evaluations-may-update.

[74] UK AI Safety Institute. "Inspect," April 21, 2024. https://inspect.ai-safety-institute.org.uk/.

[75] Japan AI Safety Institute. "AI Safety Institute." 2024. https://aisi.go.jp/wp-content/uploads/2024/07/20240627_AboutAISI_en.pdf.

## b. Setting standards

While the process for developing evaluative benchmarks, protocols, and tests that can feed into standards has been a core element of AISIs, **effectively setting and maintaining official standards is a different step[76] that only a few AISIs have plans to undertake**.

**Among the jurisdictions with AISI-like institutions, the ones with the most concrete plans related to standard-setting are the US and the EU.**

The US AISI is housed under the National Institute of Standards and Technology (NIST), which has already been publishing AI safety-relevant guidelines beyond the US AISI. For example, in 2023 NIST published their AI Risk Management Framework (RMF) aimed at providing a "flexible, structured and measurable process" to improve the trustworthiness of AI.[77] The RMF has had an influence in some jurisdictions, such as in Japan, which harmonized its AI guidelines for business to be in tune with the RMF.[78]

In 2024, NIST has published a few documents with plans for AI safety-relevant standardization:

- NIST 500-1, the Plan for Global Engagement on AI Standards
- NIST 600-1, the Artificial Intelligence Risk Management Framework: Generative Artificial Intelligence Profile
- NIST 800-1 providing guidelines on Managing Misuse Risk for Dual-Use Foundation Models (published specifically by the US AISI)

Despite this stage of NIST and US AISI's efforts being only the beginning of the standardization process (and therefore not really setting standards yet), they have explicit goals to help set standards domestically and influence the international standard-setting process. For instance, the opening paragraph of NIST 500-1 shows how these publications have been part of a longer process to help not only develop guidelines but also implement standards domestically and internationally (emphasis added):[79]

---

[76] IEEE Standards Association. "Maintaining the Standard," 2024. https://standards.ieee.org/develop/maintaining-standard/.

[77] NIST. "NIST Risk Management Framework Aims to Improve Trustworthiness of Artificial Intelligence," January 26, 2023. https://www.nist.gov/news-events/news/2023/01/nist-risk-management-framework-aims-improve-trustworthiness-artificial.

[78] Japan AI Safety Institute. "Crosswalk between the AI Operator Guidelines and the U.S. NIST AI Risk Management Framework (RMF)," April 29, 2024. https://aisi.go.jp/framework/.

> "Recognizing the importance of technical standards in shaping development and use of Artificial Intelligence (AI), the President's October 2023 Executive Order on the Safe, Secure, and Trustworthy Development and Use of Artificial Intelligence (EO 14110) calls for 'a coordinated effort…to drive the development and implementation of AI-related consensus standards, cooperation and coordination, and information sharing' internationally. Specifically, the EO tasks the Secretary of Commerce to 'establish a plan for global engagement on promoting and developing AI standards… guided by principles set out in the NIST AI Risk Management Framework and United States Government National Standards Strategy for Critical and Emerging Technology' (NSSCET). This plan, prepared with broad public and private sector input, fulfills the EO's mandate."

**The EU standard-setting efforts are closely related to the broader EU AI Act implementation process**. The Act establishes a series of obligations that general purpose AI model providers need to follow with a three-year period of gradual phase-in. During this transition period, the EU will develop standards and incentives for companies to voluntarily comply with forthcoming obligations, setting the stage to have everything ready by the time compliance by industry is required.

The Codes of Practice will be the driver of these voluntary commitments: a series of standardized guidelines and commitments that will be prepared with wide consultation from various stakeholders to which companies can adhere. Even though they are not legally binding at this stage, the EU will assess presumption of compliance based on them and, in case companies are not following them, they will have to prove compliance through other, more burdensome, means.[80]

Therefore, **the EU AI Office will drive the establishment of standards as an intermediary stage** between voluntary compliance and regulation through the AI Act. In this case, they are referred to as "harmonized standards" because they are explicitly geared towards supporting EU legislation. The bodies involved in this particular type of standard-setting include a range of organizations:[81]

- European Commission

---

[79] NIST. "A Plan for Global Engagement on AI Standards." 2024. https://doi.org/10.6028/NIST.AI.100-5, p. 1.

[80] Farrell, Jimmy. "An Introduction to Codes of Practice for the AI Act." EU Artificial Intelligence Act (blog), July 3, 2024. https://artificialintelligenceact.eu/introduction-to-codes-of-practice/.

[81] Hadrien Pouget. "Standard Setting." European Artificial Intelligence Act, 2024. https://artificialintelligenceact.eu/standard-setting/.

- European Standards Organizations (CEN, CENELEC, ETSI)
- National Standards Bodies
- European Stakeholder Organisations (e.g., Small and Medium-sized Enterprises [SMEs], trade unions, the environment, and consumers)
- Harmonized Standards Consultants (the Commission has hired EY)

The obligations covered by the AI Act that will go through the harmonized standards procedure include:[82]

- Provision of technical documentation to the AI Office and National Competent Authorities
- Provision of relevant information to downstream providers that seek to integrate their model into their AI or GPAI system (e.g., capabilities and limitations)
- Summaries of training data used
- Policies for complying with existing Union copyright law
- State of the art model evaluations
- Risk assessment and mitigation
- Serious incident reporting, including corrective measures
- Adequate cybersecurity protection

**This process makes the EU AI Office unique among AISI-like institutions: it is the only one so far whose actions are explicitly connected to a regulatory process**. All other AISIs have stayed clear of regulations to protect their technical focus and minimize political complexities.[83] But each AISI also has particularities that emerge due to their comparative advantage and limitations. In the EU, being a product of the AI Act, it is natural that the AI Office would be closely linked to regulation. It does not mean that this should be a model to other AISIs—in fact, it is still quite likely that in most jurisdictions staying clear of regulatory responsibilities is a better move for AISIs to focus on their core mission.

**The Japan AISI has also done some work that may qualify as standards-setting, though less ambitious in scope than the US and the EU**. It has created AI guidelines for business,

---

[82] Farrell, Jimmy. "An Introduction to Codes of Practice for the AI Act." EU Artificial Intelligence Act (blog), July 3, 2024. https://artificialintelligenceact.eu/introduction-to-codes-of-practice/

[83] As we explored in the Fundamental Characteristics section.

established a crosswalk between those guidelines and the NIST AI Risk Framework,[84] and started a partnership with the EU for further cooperation in business related to AI (among other technologies).[85]

On the other hand, **so far in the UK standards-setting has been left to the AI Standards Hub, led by the Alan Turing Institute**.[86] UK AISI has focused on developing standards and boosting cooperation around them, such as via the AI Safety Summits, partnerships with other AISIs, and the International Scientific Report on the Safety of Advanced AI.[87]

**Countries without AISIs have also done work on AI safety standards**. A key example is that TC260, a Chinese standardization group, published technical guidance for testing generative AI outputs for various safety or security risks.[88] Risks in scope include some that would be widely recognized elsewhere, such as ethnic discrimination or disclosure of trade secrets. Others are more specific to China, such as "incitement to [...] overthrow the socialist system." Companies also have to "pay close attention to the long-term risks that generative AI might pose."[89] Stated examples include AI deceiving humans, model self-replication, and biological weapons production.

---

[84] Japan AI Safety Institute. "AI Safety Institute." 2024.
https://aisi.go.jp/wp-content/uploads/2024/07/20240627_AboutAISI_en.pdf.

[85] European Commission. "EU and Japan Advance Joint Work on Digital Identity," April 30, 2024.
https://ec.europa.eu/commission/presscorner/detail/en/ip_24_2371.

[86] Hadrien Pouget. "What Will the Role of Standards Be in AI Governance?" Ada Lovelace Institute, April 5, 2023.
https://www.adalovelaceinstitute.org/blog/role-of-standards-in-ai-governance/.

[87] UK AI Safety Institute. "About the AI Safety Institute," Accessed September 27, 2024.
https://www.aisi.gov.uk/about.

[88] For a translation of the TC260 standard, see National Technical Committee 260 on Cybersecurity of Standardization Administration of China. "Basic Safety Requirements for Generative Artificial Intelligence Services." *Center for Security and Emerging Technology* (blog), April 4, 2024.
https://cset.georgetown.edu/publication/china-safety-requirements-for-generative-ai-final/. Other significant documents from TC260 include a draft standard on the information security of AI systems: National Information Security Standardization Technical Committee. "Information Security Technology-Security Specification and Assessment Methods for Machine Learning Algorithms." *Center for Security and Emerging Technology* (blog), February 28, 2023. https://cset.georgetown.edu/publication/china-ml-algorithm-security-draft-standard/.

and an AI safety governance framework: National Technical Committee 260 on Cybersecurity of Standardization Administration of China. "AI Safety Governance Framework," 2024.
https://www.tc260.org.cn/upload/2024-09-09/1725849192841090989.pdf.

[89] No specific measures are mandated for these risks.

## 3.4 Cooperation

The final core function of AISIs has been cooperation: the role of acting as a bridge between various groups (e.g., governments, industry, civil society, academia, other AISIs) to advance AI safety techniques, practices, and policies. In the table below, we provide an overview of cooperation activities listed among the priorities of various AISIs:

### Table 4. Objective or goal by AISIs related to cooperation

| Jurisdiction | Objective or goal related to cooperation (emphasis added) |
|---|---|
| UK[90] | ● "We are working to:<br>　○ "Test advanced AI systems and inform policymakers about their risks;<br>　○ **"Foster collaboration across companies, governments, and the wider research community to mitigate risks and advance publicly beneficial research; and**<br>　○ **"Strengthen AI development practices and policy globally."** |
| US[91] | "The AISI will focus on three key goals:<br>1. "Advance the science of AI safety;<br>2. "Articulate, demonstrate, and disseminate the practices of AI safety; and<br>3. **"Support institutions, communities, and coordination around AI safety."** |
| Japan[92] | ● "As a hub for AI safety in Japan, AISI will consolidate the latest information in industry and academia, and promote collaboration among related companies and organizations." |

Concrete cooperation activities have taken the shape of two broad categories:
- International coordination
- Scientific consensus-building

### a. International coordination

Arguably, the main efforts for international coordination have been:

- The **series of AI Safety Summits**, launched at the UK Summit in November 2023

---

[90] UK AI Safety Institute. "About the AI Safety Institute." Accessed September 27, 2024. https://www.aisi.gov.uk/about.

[91] US AI Safety Institute. "Strategic Vision." NIST, May 20, 2024. https://www.nist.gov/aisi/strategic-vision.

[92] Japan AI Safety Institute. "AI Safety Institute." 2024. https://aisi.go.jp/wp-content/uploads/2024/07/20240627_AboutAISI_en.pdf

- The **International Network of AI Safety Institutes**, launched at the Korea Summit in May 2024
- The **State of the Science Report**, started at Bletchley in November 2023, with an initial draft presented at Seoul in May 2024, and its final version due to be presented at the Paris Summit in February 2025

As with AISIs themselves, these efforts were initiated by the UK but have become jointly-led efforts as other countries have stepped up.

**The summits have become one of the main milestones for major announcements related to government and industry efforts on AI safety**, especially those touching on AISIs. In the first summit, held at Bletchley Park, the UK and the US launched their AISIs and seven top AGI companies shared detailed policies across nine AI safety policy areas.[93] In the second one, in Seoul, the international network of AISIs was announced along with the commitment of a handful of other countries (Australia, South Korea, Italy, Germany, France) to contribute to the development of this network and accelerate the advancement of the science of AI safety.

The AI Action Summit in France in February 2025 is the next major event in this series, along with a smaller convening of the international network of AI Safety Institutes in November 2024 in San Francisco. Even for the smaller convening, there is an expectation that relevant updates will be announced, such as more information about US AISI's plans and potential expansions of the international network of AISIs.[94] But the summit in France is likely to be among the most transformative ones—France has taken a different approach by deemphasizing safety in favor of innovation and "action" which may be a concerning detour from the summit's original mission. In any case, by allowing diverse perspectives on AI safety, the series of summits has managed to actively engage a varied set of stakeholders and remain a central element of the efforts on international coordination for AI safety.

---

[93] UK Government. "Policy Updates." AISS 2023, September 19, 2023. https://www.aisafetysummit.gov.uk/policy-updates/.

[94] For example, the US Commerce Department has written that "The convening will bring together technical experts on artificial intelligence from each member's AI safety institute, or equivalent government-backed scientific office, in order to align on priority work areas for the Network and begin advancing global collaboration and knowledge sharing on AI safety." US Department of Commerce. "U.S. Secretary of Commerce Raimondo and U.S. Secretary of State Blinken Announce Inaugural Convening of International Network of AI Safety Institutes in San Francisco." U.S. Department of Commerce, September 18, 2024. https://www.commerce.gov/news/press-releases/2024/09/us-secretary-commerce-raimondo-and-us-secretary-state-blinken-announce.

Intimately connected to the summits, **arguably the main concrete effort for international coordination has been the establishment of an international network of AISIs**. The network's goal is to "build 'complementarity and interoperability' between their technical work and approach to AI safety, to promote the safe, secure and trustworthy development of AI."[95] It was established by the Seoul Declaration for safe, innovative, and inclusive AI[96] signed by Australia, Canada, the EU, France, Germany, Italy, Japan, South Korea, Singapore, the US, and the UK.

## b. Scientific consensus-building

A narrower form of cooperation that seems particularly promising for AISIs is scientific consensus-building, or **the effort to build a shared understanding of AI safety as a scientific field**.

**The leader in this activity has been the UK, especially through the State of the Science Report**.[97] The report was commissioned by 30 jurisdictions represented at the first AI Safety Summit in Bletchley in November 2023, was chaired by Yoshua Bengio, and had its interim version published in May 2024. The final version will be published at the Paris Summit in February 2025. The original goal of the report as articulated in November was:

> "Rather than producing new material, it will summarise the best of existing research and identify areas of research priority, providing a synthesis of the existing knowledge of frontier AI risks. It will not make policy or regulatory recommendations but will instead help to inform both international and domestic policy making."

While this first effort was led and executed by the UK AISI, **the question of how this scientific consensus-building will continue while maintaining wide representation remains open**.

---

[95] UK Government. "Global Leaders Agree to Launch First International Network of AI Safety Institutes to Boost Cooperation of AI," May 21, 2024.
https://www.gov.uk/government/news/global-leaders-agree-to-launch-first-international-network-of-ai-safety-institutes-to-boost-understanding-of-ai.

[96] UK Government. "Seoul Declaration for Safe, Innovative and Inclusive AI by Participants Attending the Leaders' Session: AI Seoul Summit, 21 May 2024," May 21, 2024.
https://www.gov.uk/government/publications/seoul-declaration-for-safe-innovative-and-inclusive-ai-ai-seoul-summit-2024/seoul-declaration-for-safe-innovative-and-inclusive-ai-by-participants-attending-the-leaders-session-ai-seoul-summit-21-may-2024.

[97] UK Government. "'State of the Science' Report to Understand Capabilities and Risks of Frontier AI: Statement by the Chair, 2 November 2023," November 2, 2023.
https://www.gov.uk/government/publications/ai-safety-summit-2023-chairs-statement-state-of-the-science-2-november/state-of-the-science-report-to-understand-capabilities-and-risks-of-frontier-ai-statement-by-the-chair-2-november-2023.

Some proposed to keep the report under the international network of AISIs.[98] But the challenge of ensuring legitimacy and avoiding redundancy with other on-going consensus-building efforts, such as those led by the UN High-Level Advisory Board and the Organisation for Economic Co-operation and Development (OECD), remains. A potential solution could be to combine some of these efforts under the network of AISIs and a widely-representative institution, like the UN. This seems like one of the most promising ways forward to leverage the technical focus and expertise of AISIs while ensuring legitimacy.[99]

---

[98] Ziosi, Marta, Claire Dennis, Robert Trager, Simeon Campos, Ben Bucknall, Charles Martinet, Adam Smith, and Merlin Stein. "AISIs' Roles in Domestic and International Governance." Oxford Martin School, July 12, 2024. https://www.oxfordmartin.ox.ac.uk/publications/aisis-roles-in-domestic-and-international-governance.

[99] Pouget, Hadrien, Claire Dennis, Jon Bateman, Robert Trager, Renan Araujo, Haydn Belfield, Belinda Cleeland, et al. "The Future of International Scientific Assessments of AI's Risks," August 27, 2024. https://carnegieendowment.org/research/2024/08/the-future-of-international-scientific-assessments-of-ais-risks?lang=en.

# 4 Challenges and limitations

First-wave AISIs are a promising institutional model to advance and strengthen AI safety. However, like all institutions, they are not invulnerable to problems and limitations. Here, we will focus on three challenges:

- Specialization and trade-offs
- Redundancy with existing institutions
- Relationship with industry

## 4.1 Specialization and trade-offs

First-wave AISIs have by design focused on the safety of advanced AI models, which purposely **has left out pieces of AI safety and governance work**, or approach these pieces with what critics suggest is a suboptimal angle. As we have explored throughout this brief, a focus on AI safety has equipped AISIs well with a socially relevant scope that warrants government action, a technical focus that can be tackled through science, and a shared mission that facilitates cooperation across various stakeholders. At the same time, this might be seen as deprioritizing other AI-related issues, and the ability of a given country to remain innovative and competitive.

**Some researchers have also argued that the focus AISIs might have on safety evaluations as the main method for assessing safety might be insufficient to ensure advanced AI models are indeed safe.**[100] First, they argue methods like red teaming and benchmarking can be manipulated or gamed, and insufficiently cover risks such as bias and fairness concerns. Second, they posit that evaluating models in a vacuum is not enough to understand their impact on all the sectors, applications, and contexts in which they are deployed.

Coming from a different angle, **other critics claim that the focus on safety harms innovation and competitiveness**. This criticism comes especially from interest groups in the US, a country that typically takes a *laissez-faire* approach to regulating technology. In light of the fierce competition between the US and China in AI development,[101] measures that put up guardrails or potential limitations on the ability of industry to freely grow may reasonably be seen as undesirable hurdles. But groups in the EU have also raised concerns about the potential of the EU AI Act to harm European competitiveness and innovation.[102]

---

[100] Davies, Matt, Andrew Strait, and Michael Birtwistle. "Safety First?" Ada Lovelace Institute, May 17, 2024. https://www.adalovelaceinstitute.org/blog/safety-first/.

[101] Frederick, Kara, and Jake Denton. "The U.S., Not China, Should Take the Lead on AI." The Heritage Foundation (blog), October 11, 2023.
https://www.heritage.org/technology/commentary/the-us-not-china-should-take-the-lead-ai.

**One way of addressing these points of criticism is for AISIs to explicitly take into account these related problems in their prioritization**. While it may not be feasible to cover everything at once, AISIs could establish a clear prioritization order that shows their criteria for choosing certain areas first and their intention to address other areas next. One example of such a document is the NIST Plan for Global Engagement on AI Standards,[103] which outlines priority topics for standardization work and briefly outlines the criteria for this ordering.

## 4.2 Redundancy with existing institutions

As with any new institutional model, **AISIs are faced with the question of whether they are indeed necessary or if their functions could be more efficiently performed by already existing organizations.**

First, **so far AISIs have all been set up by high-income nations, deprioritizing multilateral fora that already represent a wide range of jurisdictions**. This poses at least two challenges: the relationship with China, and the inclusion of other developing countries outside the AI frontier. China is one of the leading countries in AI development. Even though it has engaged with the AI Safety Summits as a participant and has put in place several safety-relevant policies domestically, China has also pushed for international cooperation to primarily happen through the UN.[104] A UN-driven process would also contribute to the inclusion of other developing nations, as they already provide input on projects such as the Global Digital Compact.[105]

Second, some functions of **AISIs risk being redundant with field-specific international organizations, such as the Standards Developing Organizations**. AISIs might create redundancies if their processes go through organizations that are already resourced with the expertise and legitimacy necessary to advance accomplishments such as setting standards or conducting inter-governmental research projects. This also risks creating diffusion of responsibility and excess bureaucracy that could harm the mission of the AISI.

---

[102] Gikay, Asress Adimi. "Risks, Innovation, and Adaptability in the UK's Incrementalism versus the European Union's Comprehensive Artificial Intelligence Regulation." *International Journal of Law and Information Technology* 32, no. 1 (June 1, 2024): eaae013. https://doi.org/10.1093/ijlit/eaae013.

[103] NIST. "A Plan for Global Engagement on AI Standards," 2024. https://doi.org/10.6028/NIST.AI.100-5.

[104] Concordia AI. "State of AI Safety in China," October 2023. https://concordia-ai.com/wp-content/uploads/2023/10/State-of-AI-Safety-in-China.pdf.

[105] United Nations. "Global Digital Compact | Office of the Secretary-General's Envoy on Technology," 2024. https://www.un.org/techenvoy/global-digital-compact.

However, traditional multilateral organizations are often slow, and their pre-existing procedures may not be the best fit for AI safety considering the speed and complexity of the field. To balance this challenge, **one way forward would be to pair the value add brought by AISIs with the legitimacy and representation of such international organizations**.[106] On the other hand, a potential drawback of this approach is that it might slow down AISIs in the process, preventing any particular set of institutions from moving quickly.

A separate point is that **AISIs run by different countries may start to cover similar ground in a potentially redundant way.** Currently, each AISI mainly focuses on doing general AI evaluations in a way that is not specific to their jurisdiction. However formidable the technical challenge of AI safety—as a science—, and evaluations—as a method—are, it is plausible that increasing the number of AISIs focusing on these problems will hit diminishing returns; as an increasing number of countries establish AISIs, what should new AISIs focus on?

**A potential response is for various AISIs to specialize in a particular part of AI safety most relevant to their host nation and their international expertise**. For instance, countries that play a relevant role in the semiconductor supply chain, such as Japan and the Netherlands, could specialize in compute governance.[107] Countries that have access to AI experts, such as the UK and Canada, could help push the R&D side of the challenge. Jurisdictions with stronger regulatory appetite, such as the EU and Brazil,[108] could keep setting up guardrails on industry.

---

[106] See discussion of this in the context of the International Scientific Report on the Safety of Advanced AI in Pouget, Dennis et al. (2024), as well as our section above on scientific consensus-building. Pouget, Hadrien, Claire Dennis, Jon Bateman, Robert Trager, Renan Araujo, Haydn Belfield, Belinda Cleeland, et al. "The Future of International Scientific Assessments of AI's Risks," August 27, 2024. https://carnegieendowment.org/research/2024/08/the-future-of-international-scientific-assessments-of-ais-risks?lang=en.

[107] Sastry, Girish, Lennart Heim, Haydn Belfield, Markus Anderljung, Miles Brundage, Julian Hazell, Cullen O'Keefe, et al. "Computing Power and the Governance of Artificial Intelligence." arXiv, February 13, 2024. https://doi.org/10.48550/arXiv.2402.08797.

[108] Baig, Anas. "Brazil's New AI Law: What You Should Know." Securiti, March 20, 2024. https://securiti.ai/brazil-ai-regulation-and-law/.

Another possibility is the creation of regional AISIs, especially in regions with smaller countries,[109] drawing a parallel with the EU approach.

However, **while dividing the labor might mitigate redundancies, it may add more complexity and contradictions to the AISI ecosystem**. How will information sharing about model capabilities flow when one AISI is research-only and the other is housed within a regulator? When countries have established competitive measures against each other, how will they smoothly cooperate to cover different parts of the AI governance puzzle?

The international network of AISIs could also play a role in mitigating redundancies. **One model for the network could involve different levels of membership**, where core members have access to most information and commit to certain limitations on their reach, and outer levels of membership have gradually less access but more freedom in their institutional approach towards AI safety. While an institutional silver bullet is not realistic, more efficient alternatives along certain parameters, such as advancing AI safety research specifically, are possible.

Lastly, **another potential response to this criticism is that redundancy in institutions is not necessarily bad**—it often provides additional lines of defense against relevant societal issues. In the case of AISIs, even if other government agencies have AI-related expertise that could be applied to sector-specific issues, AISIs may be able to provide additional expertise or lay the groundwork for those other agencies to build up their in-house talent. Another potentially positive example would be standard-setting bodies: even if traditional organizations such as ISO and ITU are preparing to cover AI standards, AISIs may help add focused capacity that could speed up global standardization.[110]

## 4.3 Relationship with industry

As AISIs aim to understand and mitigate risks coming from cutting-edge models, it is natural that they would need to work closely with the companies developing those models. However, **AISIs need to be wary of concerns around regulatory capture**. While fears of regulatory capture are not unique to AISIs, threading this balance between AISIs having a close relationship with industry while also being prepared to be independent of industry and hold industry accountable is a challenge for governments worldwide.

---

[109] Ziosi, Marta, Claire Dennis, Robert Trager, Simeon Campos, Ben Bucknall, Charles Martinet, Adam Smith, and Merlin Stein. "AISIs' Roles in Domestic and International Governance." Oxford Martin School, July 12, 2024. https://www.oxfordmartin.ox.ac.uk/publications/aisis-roles-in-domestic-and-international-governance.

[110] Fort, Kristina. "The Role of AI Safety Institutes in Contributing to International Standards for Frontier AI Safety." arXiv.org, September 17, 2024. https://arxiv.org/abs/2409.11314v1.

Especially in the US and the UK, **AISIs have been able to establish a productive relationship with leading companies to ensure private pre-deployment access to the latest models**.[111] This collaboration is essential for AISIs to conduct safety evaluations that cover the most advanced models, particularly during the development stage, allowing governments and companies to establish guardrails before such models are made available to the public.

However, **some have argued that this close relationship may also introduce risks to AISIs' effectiveness**.[112] The worry is that the close relationship between government officials and leading companies may lead to detrimental incentives in how oversight is enforced, especially if it is based on voluntary commitments that originated from the private sector. Besides that, there is also the concern that the voluntary relationship between industry and government through AISIs may reduce the appetite for effective regulation.

To mitigate those concerns, **AISIs could prioritize transparency and keep their technical focus on research, standards, and cooperation**. While concerns of regulatory capture are relevant and should continue to be monitored, by leaving the strongest state responsibilities—such as regulation—to other government agencies, AISIs become more well-positioned to collaborate with private sector stakeholders. AISIs can then be focused on advancing the sciences of model evaluation and AI safety and advising other regulatory bodies about the relevant risks and mitigation strategies.

On the other hand, **there could be benefits to concentrating technical expertise and regulatory capacity within a single agency**. This combination could allow regulations to be more technically sound and in tune with the needs of industry. It could also facilitate the engagement of experts in civil society in academia on the regulatory process. Balancing these trade-offs is a challenge that first-wave AISIs will face as they evolve.

---

[111] NIST. "U.S. AI Safety Institute Signs Agreements Regarding AI Safety Research, Testing and Evaluation With Anthropic and OpenAI," August 29, 2024. https://www.nist.gov/news-events/news/2024/08/us-ai-safety-institute-signs-agreements-regarding-ai-safety-research and Anthropic. "Introducing Claude 3.5 Sonnet," June 21, 2024. https://www.anthropic.com/news/claude-3-5-sonnet.

[112] Stokel-Walker, Chris. "AI Needs Rules, but Who Will Get to Make Them?" Scientific American, November 3, 2023. https://www.scientificamerican.com/article/who-actually-gets-to-make-the-rules-about-ai/.

Another challenge within this category is **handling access to proprietary information in a secure and efficient way**. While companies are interested in minimizing the bureaucratic burden of different requirements across various jurisdictions, they also want to ensure that trade secrets are appropriately protected. This trade-off remains a challenge for AISIs, but their international network might be a promising venue for tackling it.

# Acknowledgements

We appreciate feedback and comments from Onni Aarne, Mauricio Baker, Christopher Covino, Romeo Dean, Oscar Delaney, Claire Dennis, George Gor, Erich Grunewald, Elliot Jones, Joe O'Brien, Alex Petropoulos, Saad Siddiqui, José Villalobos, Peter Wildeford, and Zoe Williams.

Participation in this research does not necessarily imply endorsement of this report or its findings, and the views expressed by these individuals do not necessarily reflect those of their respective organizations. All remaining errors are our own.

Leck, Andy. "Singapore: AI Verify Foundation and IMDA Launch First-of-Its-Kind Generative AI Evaluation Sandbox for Trusted AI." Baker McKenzie, November 30, 2023. https://insightplus.bakermckenzie.com/bm/data-technology/singapore-ai-verify-foundation-and-imda-launch-first-of-its-kind-generative-ai-evaluation-sandbox-for-trusted-ai.

Manancourt, Vincent. "Rishi Sunak Promised to Make AI Safe. Big Tech's Not Playing Ball. – POLITICO." Politico, April 26, 2024. https://www.politico.eu/article/rishi-sunak-ai-testing-tech-ai-safety-institute/.

Mazeika, Mantas, Long Phan, Xuwang Yin, Andy Zou, Zifan Wang, Norman Mu, Elham Sakhaee, et al. "HarmBench: A Standardized Evaluation Framework for Automated Red Teaming and Robust Refusal." arXiv.org, February 6, 2024. https://arxiv.org/abs/2402.04249v2.

Mohanty, Amlan, and Tejas Bharadwaj. "The Importance of AI Safety Institutes." Carnegie Endowment for International Peace, June 28, 2024. https://carnegieendowment.org/posts/2024/06/the-importance-of-ai-safety-institutes?lang=en.

Mokander, Jakob, Camilla Bougrine, Kevin Zandermann, Basma Abassi, and Zach Meyers. "Exploring EU-UK Collaboration on AI: A Strategic Agenda." Tony Blair Institute for Global Change, July 31, 2024. https://institute.global/insights/tech-and-digitalisation/exploring-eu-uk-collaboration-on-ai-a-strategic-agenda.

National Information Security Standardization Technical Committee. "Information Security Technology-Security Specification and Assessment Methods for Machine Learning Algorithms." *Center for Security and Emerging Technology* (blog), February 28, 2023. https://cset.georgetown.edu/publication/china-ml-algorithm-security-draft-standard/.

National Technical Committee 260 on Cybersecurity of Standardization Administration of China. "AI Safety Governance Framework," 2024. https://www.tc260.org.cn/upload/2024-09-09/1725849192841090989.pdf.

———. "Basic Safety Requirements for Generative Artificial Intelligence Services." *Center for Security and Emerging Technology* (blog), April 4, 2024. https://cset.georgetown.edu/publication/china-safety-requirements-for-generative-ai-final/.

IAPS | Institute for AI Policy and Strategy    Understanding AISIs | 44

NIST. "A Plan for Global Engagement on AI Standards." National Institute of Standards and Technology, 2024. https://doi.org/10.6028/NIST.AI.100-5.

———. "AI Test, Evaluation, Validation and Verification (TEVV)." *NIST*, July 5, 2022. https://www.nist.gov/ai-test-evaluation-validation-and-verification-tevv.

———. "Artificial Intelligence Safety Institute Consortium (AISIC)," April 15, 2024. https://www.nist.gov/artificial-intelligence-safety-institute/artificial-intelligence-safety-institute-consortium-aisic.

———. "NIST Calls for Information to Support Safe, Secure and Trustworthy Development and Use of Artificial Intelligence." *NIST*, December 19, 2023. https://www.nist.gov/news-events/news/2023/12/nist-calls-information-support-safe-secure-and-trustworthy-development-and.

———. "Strategic Vision." *NIST*, May 20, 2024. https://www.nist.gov/aisi/strategic-vision.

———. "U.S. AI Safety Institute Signs Agreements Regarding AI Safety Research, Testing and Evaluation With Anthropic and OpenAI." *NIST*, August 29, 2024. https://www.nist.gov/news-events/news/2024/08/us-ai-safety-institute-signs-agreements-regarding-ai-safety-research.

———. "U.S. Artificial Intelligence Safety Institute." *NIST*, October 26, 2023. https://www.nist.gov/aisi.

———. "U.S. Secretary of Commerce Gina Raimondo Releases Strategic Vision on AI Safety, Announces Plan for Global Cooperation Among AI Safety Institutes." *NIST*, May 21, 2024. https://www.nist.gov/news-events/news/2024/05/us-secretary-commerce-gina-raimondo-releases-strategic-vision-ai-safety.

Petropoulos, Alex. "The AI Safety Institute Network: Who, What and How?" *ICFG* (blog), September 10, 2024. https://icfg.eu/the-ai-safety-institute-network-who-what-and-how/.

Petropoulos, Alex, and Max Reddel. "The EU AI Office Needs Top Scientific Talent, Not Familiar Faces." www.euractiv.com, September 10, 2024. https://www.euractiv.com/section/digital/opinion/the-eu-ai-office-needs-top-scientific-talent-not-familiar-faces/.

Pouget, Hadrien. "Standard Setting | EU Artificial Intelligence Act." *EU Artificial Intelligence Act* (blog), March 2024. https://artificialintelligenceact.eu/standard-setting/.

Pouget, Hadrien, Claire Dennis, Jon Bateman, Robert Trager, Renan Araujo, Haydn Belfield, Belinda Cleeland, et al. "The Future of International Scientific Assessments of AI's Risks,"

———. "Grants | AISI," 2024. https://www.aisi.gov.uk/grants.

———. "Inspect." Inspect, April 21, 2024. https://inspect.ai-safety-institute.org.uk/.

———. "Research | AISI," 2024. https://www.aisi.gov.uk/category/research.

UK Government. "AI Safety Institute Approach to Evaluations." GOV.UK, February 9, 2024. https://www.gov.uk/government/publications/ai-safety-institute-approach-to-evaluations/ai-safety-institute-approach-to-evaluations.

———. "AI Safety Institute: Third Progress Report." GOV.UK, February 5, 2024. https://www.gov.uk/government/publications/uk-ai-safety-institute-third-progress-report/ai-safety-institute-third-progress-report.

———. "Chair's Summary of the AI Safety Summit 2023, Bletchley Park." GOV.UK, November 2, 2023. https://www.gov.uk/government/publications/ai-safety-summit-2023-chairs-statement-2-november/chairs-summary-of-the-ai-safety-summit-2023-bletchley-park.

———. "Frontier AI Taskforce: First Progress Report." GOV.UK, September 7, 2023. https://www.gov.uk/government/publications/frontier-ai-taskforce-first-progress-report/frontier-ai-taskforce-first-progress-report.

———. "Frontier AI Taskforce: First Progress Report." GOV.UK. Accessed August 4, 2024. https://www.gov.uk/government/publications/frontier-ai-taskforce-first-progress-report/frontier-ai-taskforce-first-progress-report.

———. "Frontier AI Taskforce: Second Progress Report." GOV.UK, October 30, 2023. https://www.gov.uk/government/publications/frontier-ai-taskforce-second-progress-report/frontier-ai-taskforce-second-progress-report.

———. "Global Leaders Agree to Launch First International Network of AI Safety Institutes to Boost Cooperation of AI." GOV.UK, May 21, 2024. https://www.gov.uk/government/news/global-leaders-agree-to-launch-first-international-network-of-ai-safety-institutes-to-boost-understanding-of-ai.

———. "Initial £100 Million for Expert Taskforce to Help UK Build and Adopt next Generation of Safe AI." GOV.UK, April 24, 2024. https://www.gov.uk/government/news/initial-100-million-for-expert-taskforce-to-help-uk-build-and-adopt-next-generation-of-safe-ai.

Understanding AISIs | 47

———. "International Scientific Report on the Safety of Advanced AI." GOV.UK, May 17, 2024. https://www.gov.uk/government/publications/international-scientific-report-on-the-safety-of-advanced-ai.

———. "Leading Frontier AI Companies Publish Safety Policies." GOV.UK. Accessed August 11, 2024. https://www.gov.uk/government/news/leading-frontier-ai-companies-publish-safety-policies.

———. "Policy Updates." AISS 2023, September 19, 2023. https://www.aisafetysummit.gov.uk/policy-updates/.

———. "Seoul Declaration for Safe, Innovative and Inclusive AI by Participants Attending the Leaders' Session: AI Seoul Summit, 21 May 2024." GOV.UK, May 21, 2024. https://www.gov.uk/government/publications/seoul-declaration-for-safe-innovative-and-inclusive-ai-ai-seoul-summit-2024/seoul-declaration-for-safe-innovative-and-inclusive-ai-by-participants-attending-the-leaders-session-ai-seoul-summit-21-may-2024.

———. "Seoul Ministerial Statement for Advancing AI Safety, Innovation and Inclusivity: AI Seoul Summit 2024." GOV.UK, May 22, 2024. https://www.gov.uk/government/publications/seoul-ministerial-statement-for-advancing-ai-safety-innovation-and-inclusivity-ai-seoul-summit-2024/seoul-ministerial-statement-for-advancing-ai-safety-innovation-and-inclusivity-ai-seoul-summit-2024.

———. "'State of the Science' Report to Understand Capabilities and Risks of Frontier AI: Statement by the Chair, 2 November 2023." GOV.UK, November 2, 2023. https://www.gov.uk/government/publications/ai-safety-summit-2023-chairs-statement-state-of-the-science-2-november/state-of-the-science-report-to-understand-capabilities-and-risks-of-frontier-ai-statement-by-the-chair-2-november-2023.

———. "The Bletchley Declaration by Countries Attending the AI Safety Summit, 1-2 November 2023." GOV.UK, November 1, 2023. https://www.gov.uk/government/publications/ai-safety-summit-2023-the-bletchley-declaration/the-bletchley-declaration-by-countries-attending-the-ai-safety-summit-1-2-november-2023.

———. "UK-Canada Science of AI Safety Partnership." GOV.UK, May 20, 2024. https://www.gov.uk/government/publications/uk-canada-science-of-ai-safety-partnership/uk-canada-science-of-ai-safety-partnership.
Understanding AISIs | 48